\documentclass[manuscript]{emulateapj}
\slugcomment{Published in the Astrophysical Journal, 705:226-236, 2009 November 1}

\usepackage{epsf,graphicx}
\shorttitle{MUSTANG and GISMO observations of M42.}
\shortauthors{Dicker et al.}

\begin{document}

\title{90\,GHz and 150\,GHz observations of the Orion M42 region. \\ A sub-millimeter to
radio analysis. }


\author{S. R. Dicker\altaffilmark{1},
B. S. Mason\altaffilmark{2},
P. M. Korngut\altaffilmark{1},
W. D. Cotton\altaffilmark{2},
M.  Compi\`egne\altaffilmark{3},
M. J. Devlin\altaffilmark{1},
P. G. Martin\altaffilmark{3},
P. A. R Ade\altaffilmark{4},
D. J. Benford\altaffilmark{5},
K. D. Irwin\altaffilmark{6},
R. J. Maddalena\altaffilmark{7},
J. P. McMullin\altaffilmark{7}, D.S. Shepherd\altaffilmark{7},
A. Sievers\altaffilmark{8},
J. G. Staguhn\altaffilmark{5,}\altaffilmark{9},
and C. Tucker\altaffilmark{4}}
\altaffiltext{1}{University of Pennsylvania, 209 S. $33^{\mbox{rd}}$ St, Philadelphia, PA 19104.}
\altaffiltext{2}{National Radio Astronomy Observatory,  Charlottesville, VA 22903.}
\altaffiltext{3}{Canadian Institute for Theoretical Astrophysics, University of Toronto, 60  St George St, Toronto, ON M5S 3H8, Canada.}
\altaffiltext{4}{Cardiff University, 5 The Parade, Cardiff, CF24 3YB, UK.}
\altaffiltext{5}{NASA Goddard Space Flight Center, Greenbelt, MD 20771.}
\altaffiltext{6}{National Institute of Standards and Technology, 325  Broadway, Boulder, CO 80305.}
\altaffiltext{7}{National Radio Astronomy Observatory, Green Bank, WV 24944.}
\altaffiltext{8}{IRAM, Avenida Divina Pastora, 7, Nucleo Central, E
  18012 Granada, Spain}
\altaffiltext{9}{University of Maryland, College Park, MD 20742.}

\begin{abstract}

  We have used the new 90\,GHz MUSTANG camera on the Robert C. Byrd Green Bank
  Telescope (GBT) to map the bright Huygens region of the star-forming
  region M42 with a resolution of 9$''$ and a sensitivity of
  2.8\,mJy\,beam$^{-1}$.  Ninety GHz is an interesting transition frequency, as
  MUSTANG detects both the free-free emission characteristic of
  the H{\footnotesize\,II} region created by the Trapezium stars,
  normally seen at lower frequencies, and thermal dust emission from
  the background OMC1 molecular cloud, normally mapped at higher
  frequencies.  We also present similar data from the 150\,GHz GISMO
  camera taken on the IRAM 30\,m telescope.  This map has 15$''$
  resolution.   By combining the MUSTANG data with 1.4,
  8, and 21\,GHz radio data from the VLA and GBT, we 
derive a new estimate of the emission
  measure (EM) averaged electron temperature of
  $T_e\,=\,11376\pm1050$\,K by an original method relating free-free
  emission intensities at optically thin and optically thick
  frequencies.  Combining {\it Infrared Space Observatory}-longwavelength spectrometer ({\it ISO}-LWS) data with our data, we derive a new
  estimate of the dust temperature and spectral emissivity index
  within the 80$''$ ISO-LWS beam toward Orion\,KL/BN,
  $T_d\,=\,42\pm3\,K$ and $\beta_d\,=\,1.3\pm0.1$.  We show that both 
  $T_d$ and $\beta_d$ decrease when going from the H{\footnotesize\,II} 
region and excited OMC1 interface to 
 the denser UV shielded part of OMC1 (Orion\,KL/BN, Orion\,S). 
With a model consisting of only free-free and thermal dust
  emission we are able to fit data taken at frequencies from 1.5\,GHz
  to 854\,GHz (350\,$\mu$m).
\end{abstract}


\section{INTRODUCTION}

The Orion Nebula (M42), located 437$\pm$19\,pc from the Sun
\citep{Hirota2007}, is one of the closest regions of active high mass
star formation. This H{\footnotesize\,II} region, which lies in front of
the OMC1 molecular cloud, is excited by a group of OB stars known as
the Trapezium. The OMC1 molecular cloud is part of a bigger 
complex that extends over 30$^\circ$ of the sky. It is an ideal site
for the study of star formation and the physics of the interstellar
medium (ISM)\@.  The M42 area has been extensively mapped across the
electromagnetic spectrum \citep[see][for a review]{Odell2001}. This
area contains  hot young stars located in the Trapezium,
pre-stellar cores, and regions of dense molecular gas.  With
multi-wavelength observations the properties of these different
regions and how they interact can be understood.

In this paper we report on some of the first scientific observations
at 90\,GHz using the 100\,m diameter Robert C. Byrd Green Bank Telescope (GBT).  A 9$''$
resolution $5' \times 9'$ continuum map in the bright Huygens region
of M42 was made using the new MUSTANG focal plane array described in
Section~\ref{MUSTANG}.  To confirm the reliability of the MUSTANG
images, two independent data analysis pipelines were used. These are
described in Sections~\ref{maxlike} and~\ref{Obit}.

At 90 GHz the Huygens region is bright, not simply with free-free
radiation from the ionized gas but also, in remarkably equal measure,
with thermal dust radiation from the molecular cloud.  However, the two
emission components have quite different spatial structure. We
were able to separate them using data at higher and lower frequencies
between 1.5\,GHz and 854\,GHz.  These data
are presented in Section~\ref{other_wavelengths}.  We describe the morphology
of OMC1-M42 as seen over this frequency range in
Section~\ref{sec:morphology}. In Section\,\ref{sect:ff_correl}, we
derive an estimate of the electron temperature and in
Section\,\ref{sect:dust}, we study the dust emission. The results are
summarized in Section~\ref{conc}.

\section{THE MUSTANG CAMERA}\label{MUSTANG}
MUSTANG, the MUltiplexed Squid TES Array at Ninety GHz \citep{SPIE2008}
 is a continuum camera built as a user instrument for
 the GBT. At the heart of the instrument is an $8\times 8$ focal plane
 array of Transition Edge Sensor (TES) bolometers built at NASA/GSFC. 
Two high density polyethylene 
lenses reimage the Gregorian focus of the GBT onto the detectors
 with an effective focal length of 162\,m so that each detector is
 spaced on the sky by $4.2''$, approximately $0.5 f \lambda
 $.\footnote{During the summer of 2008 this spacing was increased to
   $5.7''$ in order to obtain better signal to noise.}  A $4.2''$ spacing
 fully samples the sky in a single pointing of the GBT, decreasing 
 the need for fast ($>$0.1\,Hz) chopping or scanning in order 
 to reduce $1/f$ noise. Slow movements of the telescope ($\sim 1'$ per
 second) produce many redundant measurements of each point in the
 field of view which can be used to remove much of the noise on
 timescales from 0.07\,s to 0.5\,s (the time it takes a source to move
 between pixels and across the array). The scan pattern constrains
 $1/f$ noise on time scales longer than $\sim$5\,s (the time between
 observations of the same part of the sky). 
Lower scanning speeds are a great advantage as
 large accelerations can excite oscillations in the GBT's
 structure making accurate pointing problematic.

MUSTANG has an 81\,GHz to 99\,GHz bandpass. Its optical system 
 provides a uniform illumination of the primary
mirror of the GBT out to a radius of 45\,m and zero
elsewhere.  A best fit Gaussian to the measured beam shape 
has a FWHM of 9$''$, close to the value predicted by optical models of the
instrument. Below 10\,dB the measured beam is significantly higher than
that expected from a perfect telescope.  This is due to
power being scattered from small errors in the shape of the GBT's
primary mirror that occur on length scales of 1--10\,m. 
During the observations presented in this
paper, the surface had a 390\,$\mu \rm{m}$ RMS, consistent with our
measured beam efficiency of 10\%. Since these observations were made 
there have been significant improvements to the GBT's surface and our
measured beam efficiency is currently (in March 2009) around 20\%. 

The output from each detector is obtained at a software selectable
rate which for these observations was 
1\,kHz. The data values are relative to an
arbitrary but stable zero point and are in arbitrary units of 
  ``counts''.  The conversion of counts to flux units depends on each
detector's gain which varies depending on several factors including
the bias voltage and the location of the TES on its transition.  The
gain of the detectors can be measured using a small black 
body, ``CAL'', located at the Lyot stop so it
 uniformly illuminates all the detectors.   Tests have shown detector
gains to be stable to better than a few percent over many hours.

\section{OBSERVATIONS}

The MUSTANG map is the result of 5.6 hours of 
integration time yielding a final map RMS of 2.8\,mJy\,beam$^{-1}$.
Observations were conducted over 4 sessions in January
and February 2008. Each observing session was begun by mapping
the primary calibrators Mars or Saturn. In this paper we use the
\citet{Ulich1981} measurement for the 90\,GHz
brightness temperature of Saturn, $T_B = 149.3$\,K. 
Every 30 minutes focus
and pointing corrections were measured and applied by collecting small maps of a
bright source at a range of focus settings. In cases where these data
indicated that the pointing correction was significantly less than
MUSTANG's instantaneous field of view, corrections were applied in the
data analysis offline.   The bright quasar 0530+135 was used for this purpose as it
is unresolved and is located only several degrees away from
M42. The gains of the detectors were measured by taking a
30 second scan while pulsing CAL at 0.5\,Hz. Another 30 second
scan on a blank piece of sky was taken to aid with estimations 
of the weights used when co-adding data from different detectors.

When scanned across the sky a fully-sampled imaging detector array
such as MUSTANG produces many redundant measurements of a given
Nyquist sky pixel. In addition to the sky signal, systematic signals
(e.g., atmospheric emission
and gain drifts) are present. For a well-designed system these will change
slowly compared to the sky signal and have a different characteristic
signature in the detector time stream. This is facilitated by an
appropriate sky modulation (scan) pattern. Furthermore, most of the
dominant systematic signals --- in particular atmospheric emission and
thermal drifts in the receiver --- are common-mode between
detectors.  Provided that the sky map is highly oversampled on most
spatial scales it is, in principle, straightforward to form
high-quality astronomical images.

\begin{figure}[t!]
 \centering
 \includegraphics[width=0.45\textwidth]{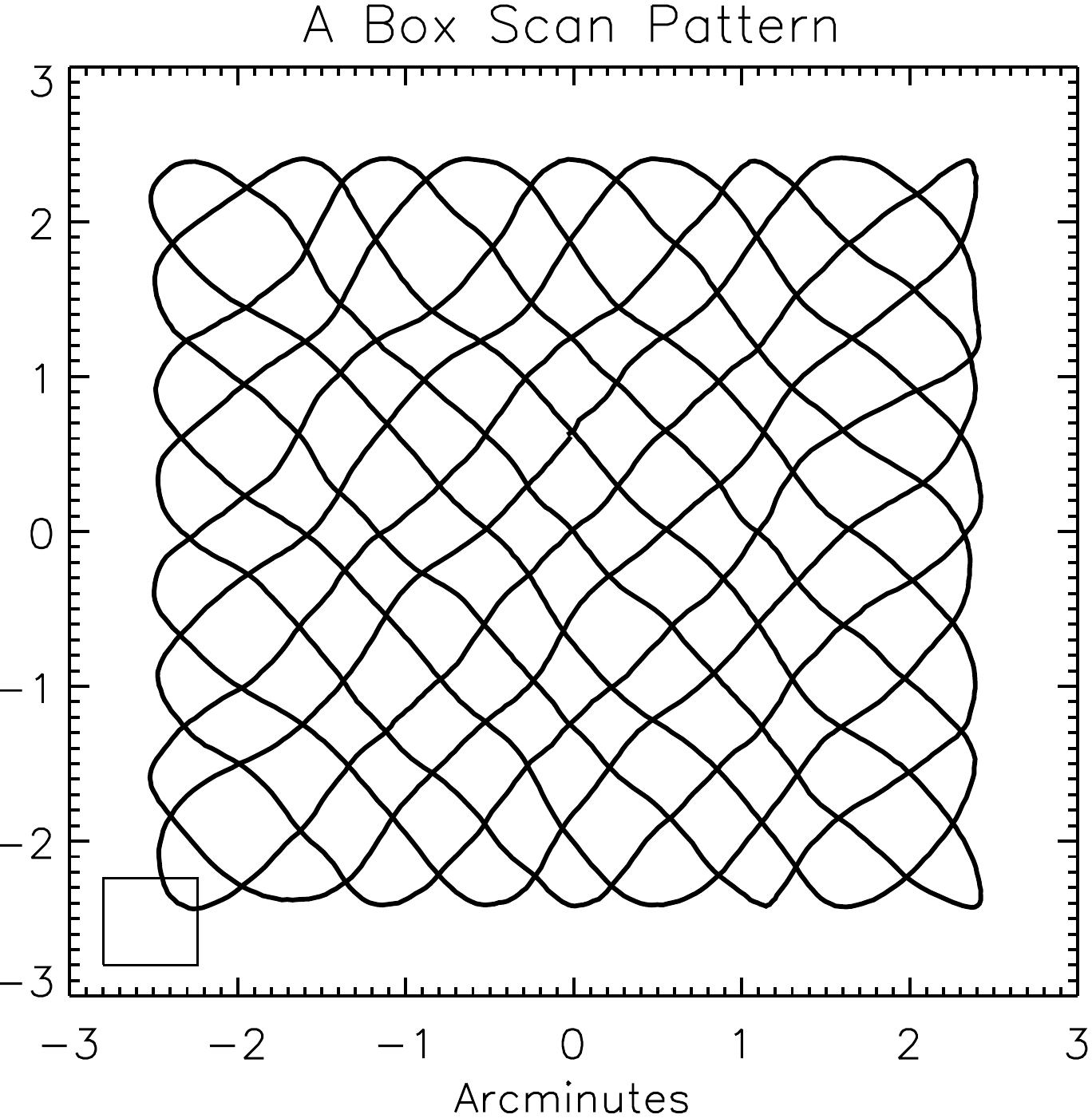}
 \caption{The GBT boresight trajectory on the sky during
   the ``box'' scan pattern used in our observations.  The
   square in the bottom left represents MUSTANG's 33.6$''$ square field of view.}
 \label{fig:scan_pattern}
 \end{figure}

M42 was mapped primarily with tiled $5' \times 5'$ ``box'' scans which
take 5 minutes each.  The telescope is scanned
diagonally across the rectangular area being mapped at a constant speed,
turning around at the edges with a gentle acceleration so as to
minimize  resonances in the GBT's structure
(Figure~\ref{fig:scan_pattern}). The scan speed is chosen so that all
astronomical objects with spatial scales smaller than a few arcminutes
pass through the beam in under 5 seconds, a timescale faster than
changes in the atmosphere or instrumental response.  
This scan pattern generates uniform coverage with a high degree of
cross linking on many different timescales enabling low frequency
instrumental and atmospheric signals to be removed.  The
$5' \times 9'$ selected region of M42 was mapped in sets of 6 box
scans on subdivided north and south fields.  To provide additional
cross linking, the central pointing of each scan was offset from the
previous one by a non-integral multiple of the spacing between crossings.

\begin{figure*}[t!]
   \centering
    \includegraphics[width=1.0\textwidth]{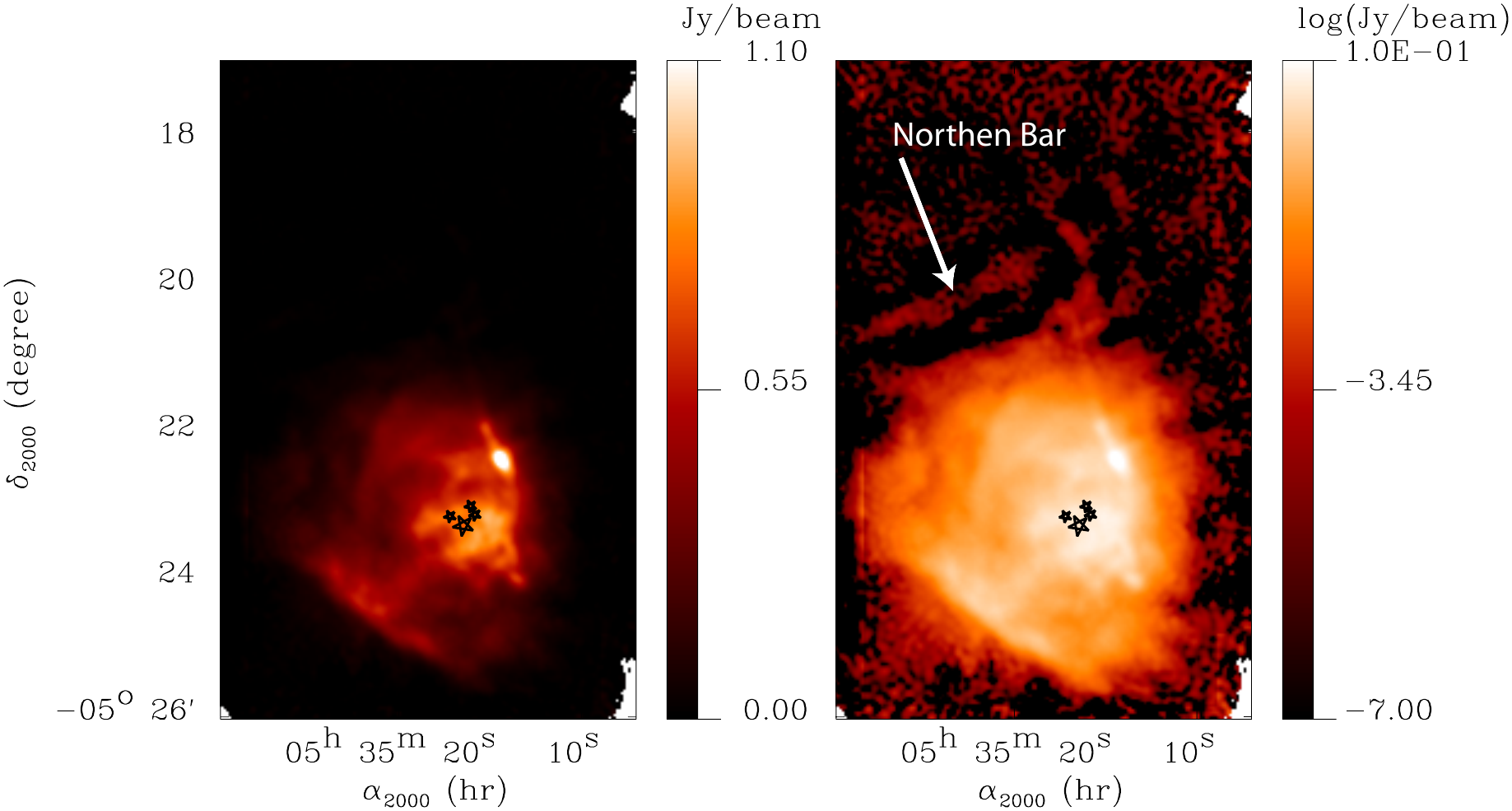}
       \caption{The M42 MUSTANG map reduced with Obit
         (Section~\ref{Obit}).  The large dynamic range, over a factor
         of 1000, can be seen by
         comparing the linear scale (left) and the
         logarithmic scale (right). 
         Shown on each map are the Trapezium stars with the bigger one being $\Theta^1$\,C\,Ori, the main ionization source.}
         \label{fig:imaffi_mustang}
\end{figure*}

\section{MUSTANG DATA REDUCTION}

MUSTANG's data were reduced using two independent
pipelines.  The first, written in 
IDL\footnote{http://www.ittvis.com/ProductServices/IDL.aspx}, 
implements a least-squares map-making
approach explicitly solving for instrumental parameters and sky
pixel values.  The second, known as 
Obit\footnote{http://www.cv.nrao.edu/~bcotton/Obit.html}
\citep{cotton08}, is an iterative algorithm
that uses a single-dish implementation of CLEAN. The following subsections
describe each approach.  Several general properties of the MUSTANG
data are worth noting:
\begin{enumerate}

\item The largest signal component is due to atmospheric emission.
  Due to the large size of the GBT and the small divergence of the
  beams, each detector samples the same volume of atmosphere.  Once the
  detectors are properly calibrated this signal is almost completely
  common mode.

\item  There is also a strong component due to MUSTANG's pulse tube
cooler causing temperature fluctuations in the optical path. This  
signal is nearly 
perfectly sinusoidal and its frequency is very stable at 
$1.4117\pm0.005 \, {\rm
Hz}$.  The best way to remove this signal was found to be by fitting a
sine wave with a slowly varying amplitude to the
low pass filtered time streams.

\item There are significant ($0.1-0.2$ second) absolute timing offsets
in the MUSTANG data.  At our typical slew speeds of $1'$ per second
 this corresponds to pointing offsets in the scan direction comparable
to the MUSTANG beam. A 1 pulse-per-second (1PPS) signal is available
in the GBT receiver room which demarks UTC second boundaries. This is
sampled synchronously with the detectors and used to correct
the instrument's timestamps offline.

\end{enumerate}

Since most of the systematic effects are common mode and a naive
common mode subtraction will remove any sky structures larger than the
instantaneous field of view, we wanted to carefully test the efficiency
of our imaging algorithms at retrieving the large scale emission.
Consequently we analyzed the data with two independent approaches
and compared the resulting maps, as described in the following sections.

\subsection{Least Squares Map Making}\label{maxlike}

The map-making problem can be framed as a large linear least
squares problem. As an example consider that we have $N_{det}$
detectors, each yielding $N_{int}$ integrations for a total of
$N_{tot} = N_{det} \times N_{int}$ data points. 
The time stream data can be represented by a single data vector
$\Delta$ with $N_{tot}$ components.  The observation covers a map
comprising $N_{pix} \ll N_{tot}$ sky pixels. Furthermore we allow $N_{par}
\ll N_{tot}$ parameters to describe instrumental systematics --- for
instance, a slowly varying common mode due to sky emission and a set
of fixed detector gains.  The sky plus instrument model parameters are represented by a
vector, $\delta$, with $N_{pix}+N_{par}$ parameters.  If the instrument
model is linear, or linearizable, the data are related to the model
parameters by a $N_{tot} \times (N_{pix}+N_{par})$ matrix $H$ by the
equation
\begin{equation}
\Delta = H \delta
\end{equation}
As long as $N_{tot} > N_{pix} + N_{par}$, and $H$ is non-singular,
a solution of the form
\begin{equation}
\label{eq:lsq}
\delta = (HWH)^{-1} \, HW \, \Delta .
\end{equation}
exists where we have introduced a weight matrix, $W$, describing the noise and
correlations in the time stream data. An advantage of this approach is that it
is possible to explicitly determine the uncertainties in the resulting
instrument and sky model parameters as well as their correlations. 
This approach has been widely used for imaging the
Microwave Background and in other deep extragalactic experiments 
\citep[e.g., ][]{tegmark, patanchon}.  Two difficulties are
present. First, the matrix $HWH$ is of size $(N_{pix} + N_{par})^2$,
which can be unwieldy if treated with brute force. 
Second, if we desire to
solve for detector gains, the problem is no longer linear in the model
parameters, but rather is bilinear due to ${\rm gain} \times {\rm sky
\, pixel}$ terms.

\citet{fmh00} have implemented an iterative solution to
Eq.~\ref{eq:lsq} which, for typical data sets, can be executed quickly
on modern desktop computers. We have integrated their least squares
imaging routines into an IDL data analysis pipeline initially
developed by the project team for quick-look and exploratory data
reduction. Several models are provided by the \citet{fmh00} code.
We have explored variants of a model of the form
\begin{equation}
\Delta = G ( S + A ) + F,
\end{equation}
where $G$ is a set of detector gains, $S$ is the sky (astronomy)
model, $A$ is a common mode optical loading term, and $F$ is a
per-detector additive offset. Care must be taken that the number of
model parameters do not become excessive, which can be done by a
combination of selecting a long solution interval for the variable
terms, and by fixing or eliminating degenerate terms. For the MUSTANG
data presented in this paper
a comprehensive noise characterization was not necessary.
The weight matrix $W$ was assumed to be diagonal with per-detector
variances determined every half hour from the time stream model
residuals.  Imaging fainter extended sources will benefit from a more
accurate treatment.

The final least squares map was produced by imaging roughly half-hour long sets of
scans and coadding the resulting images, properly accounting for the
pixel weight distributions.  The $G \times A$ models
have an exact degeneracy which corresponds to an arbitrarily oriented
two-dimensional linear ramp in pixel intensities across the map. This
was eliminated by fitting a plane to two orthogonal, approximately
emission-free lines of pixels in each half-hour map and subtracting it
before coaddition.

\subsection{The iterative CLEAN algorithm --- Obit}\label{Obit}

The basic approach used in the Obit reduction of MUSTANG data is to
iteratively estimate the background signal and the astronomical
signal (the ``sky model'').  At the start of each iteration loop, the current 
sky model is subtracted from the time stream data and  the background
signals (atmosphere, detector, etc.) are estimated by low-pass filtering
the residual.  These backgrounds are then removed from the original
time streams, an observed sky image is produced, and the true sky
estimated using CLEAN\@.  The sky model and the backgrounds are iteratively
refined by reducing the timescale of the low-pass filter used in the
background estimation as the iteration progresses. 

Time stream data are imaged by multiplying each datum with a
griding kernel producing a continuous function which is then sampled
onto a regular celestial grid.  Two grids are accumulated. The first
grid, the image, contains the data
samples multiplied by a data weight and the griding function. The
second grid contains
the data weights multiplied by the griding function.  When all data are
accumulated, the image is normalized by dividing by the second grid 
on a pixel-by-pixel basis.
The griding kernel was chosen to minimize the broadening of the
instrumental response and is an exponential times a sinc function.

There are two types of operations used by Obit, 
those done independently on each 5 minute scan and
those which are coupled to all data through the sky model.  The
independent processing steps are the following.

\begin{enumerate}

\item {\it Reading archive data.}  The kHz sampled data are
  read and averaged to a 20 Hz rate.  Then the 1.4117\,Hz signal
   from the pulse tube refrigerator is estimated and removed by
   fitting a sine wave with a common phase and a detector-dependent
   amplitude.

\item {\it Gain calibration.}  Detector gains are determined from the half
  hourly CAL 
   scans.  These gains are applied to all subsequent data until the next
   calibration scan.  Calibration to Jansky is achieved from
   observations of planets. 

\item {\it Weight calibration.}
   Statistical weights are determined for each detector during blank
   sky observations
   by the RMS residuals following the fitting and removal of a low
   order polynomial from the data.  Dead and badly behaving
   detectors are given zero weight.


\item {\it Common atmosphere and detector offsets.}
   A single detector offset per scan and a time variable common
   ``atmospheric'' term is estimated by low pass filtering the
   median of all working detectors in each time sample.

\end{enumerate}

When these operations are complete, multiple scans over multiple days
can be jointly imaged and non-astronomical backgrounds removed.  The
imaging/background removal cycle consists of the following.

\begin{enumerate}

\item {\it Imaging.}
   The current estimates of the backgrounds are removed and all data
   are accumulated onto a common grid and normalized to form the
   ``Raw'' image. 

\item {\it Deconvolution.}
   CLEAN is used to generate a sky model consisting of delta
   functions. Fixed or interactive specification of the CLEAN windows
   is used to guide the process.

 \item {\it Subtracting the model.}  The CLEAN delta functions are
   convolved with a Gaussian approximation to the telescope beam
   shape. The value corresponding to each data sample is interpolated
   from this CLEAN image and subtracted from the time stream data.
   This results in a residual time stream in which the current
   estimate of the astronomical signal has been removed.

\item {\it Estimation of the background.}
   An estimation of the background signal is made by low pass
   filtering of the residual time stream data.  This filtering
   alternates between estimating individual detector offsets and a
   common mode offset estimated with shortest time scales one third of
   that used for the detector offsets.
\end{enumerate}

After the cycles have converged, the final CLEAN image is used for
subsequent analysis.

\subsection{Comparison of Results}

The map produced by the Obit pipeline is shown in
Figure~\ref{fig:imaffi_mustang}.  On a
linear scale the Obit maps are very similar to those produced by least
squares.  The peak height of Orion KL, the brightest feature in the
maps was within 5\% of each other 
and neither data pipeline produced significant 
 negative flux. For bright features, all extended structure on 
angular scales up to 2$'$ was recovered.  
The most prominent discrepancies between
 the maps were faint
extended structures in the north of the maps 
which have higher signal-to-noise in the Obit maps.
Given the similarity of the
maps produced by independent analysis we believe all of the features
above the stated noise level to be real.

\section{ORION AT OTHER FREQUENCIES}\label{other_wavelengths}

\begin{table*}[!t]
  \centering
  \caption{A summary of the different frequency maps used in this
    paper.}
  \label{tab:datasets} 
  \begin{tabular}{l c c c c c c c}
   \hline           
   \hline           
          Map     & $\lambda$  & $\nu$ &  FWHM  & largest Angular &  Relative       &  RMS &  RMS\\
               &   (mm)     & (GHz) & ($''$) & Scale ($'$) & uncertainty    &  (MJy\,sr$^{-1}$) & (mJy beam$^{-1}$) \\
   \hline
   SHARC-II 350 &  0.35     & 854  &   9     & $>2$ & 30\% &  509 & 1100 \\
   SCUBA 450    &  0.45     & 664  &  8$^a$  & 1.1  & 30\% &  106 &  300  \\
   SCUBA 850    &  0.85     & 352  &  15$^a$ & 1.1  & 15\% &   5.8 &  40  \\
   GISMO        &  2.0      & 150  & 15      & $>2$ & 20\% &   17 &  100  \\
   MUSTANG      &  3.3      & 90   & 9       & 2    & 15\% &   1.3  &   2.8 \\
   $K$ band       &  14       & 21.5 & 33.5    & 6    & 15\% &  6  &   180 \\
   $X$ band       &  35       & 8.4  & 8.4     & 2    & 15\% &   3.7  &   7 \\
   $L$ band       &  200      & 1.5  & 7       & 14   & 15\% &   2.8  &  3.6 \\
   \hline
  \end{tabular}

$^a$ The SCUBA beam shapes consist of a core Gaussian with a FWHM given in
  this table and a broad component with a relative peak intensity of
  5\% of the core and a FWHM of 30$''$ \citep{Johnstone2006}.  
\end{table*}

\begin{figure*}[!t]
   \centering
    \includegraphics[width=1.0\textwidth]{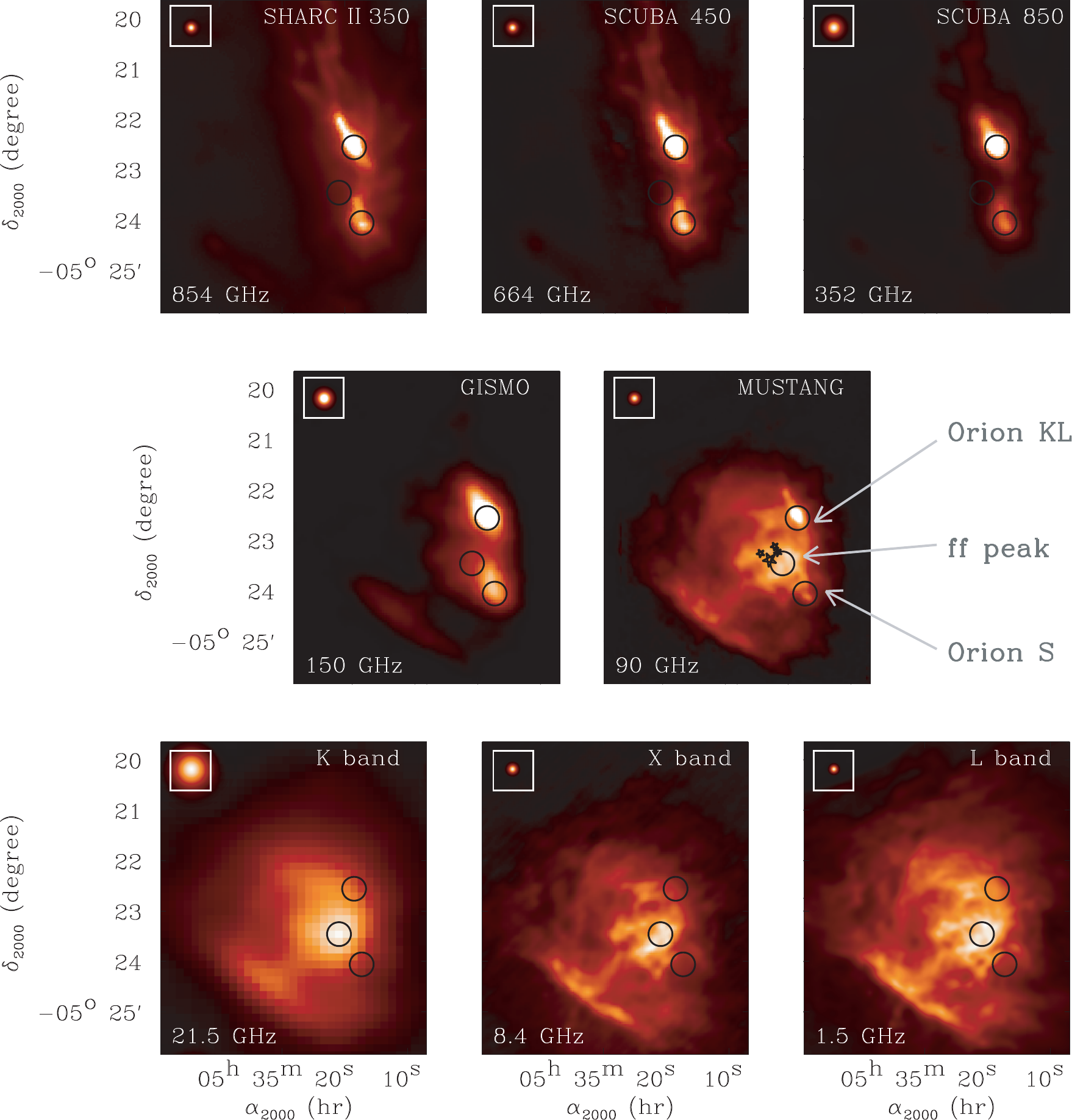}
       \caption{Maps of M42. All datasets are displayed with their nominal 
                resolution.  The inset boxes in each map 
                display Gaussian approximations
                to the point-spread-functions.  The 30$''$ circles on each map
                delineate areas used to compute average spectra
                at Orion KL, Orion\,S and the free-free emission
                peak locations (shown in
                Figure\,\ref{fig:spectra}).
                 Shown on the MUSTANG map are the Trapezium stars with the bigger one being $\Theta^1$\,C\,Ori, the main ionization source.}
         \label{fig:imaffi_all}
\end{figure*}

\begin{figure*}[!t]
   \centering
    \includegraphics[width=1.0\textwidth]{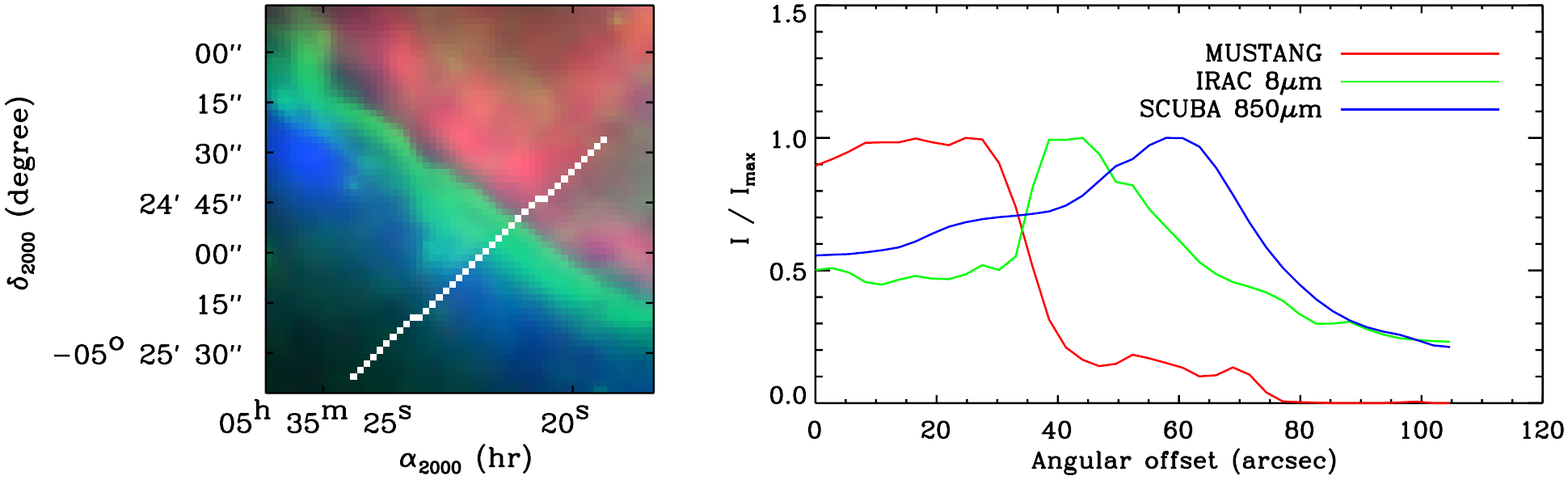} 
    \caption{A false color map of the Orion bar. The SCUBA 850\,$\mu$m
      image is shown in blue, the IRAC $8\,\mu$m map is shown in
      green, and red is the MUSTANG data.  The profiles displayed
      correspond to the cut shown on the map, starting from the upper
      right and moving away from the Trapezium stars which are located
      toward the northwest of this map.}
         \label{fig:OrionBar}
\end{figure*}

\begin{figure}[!t]
   \centering
    \includegraphics[width=0.45\textwidth]{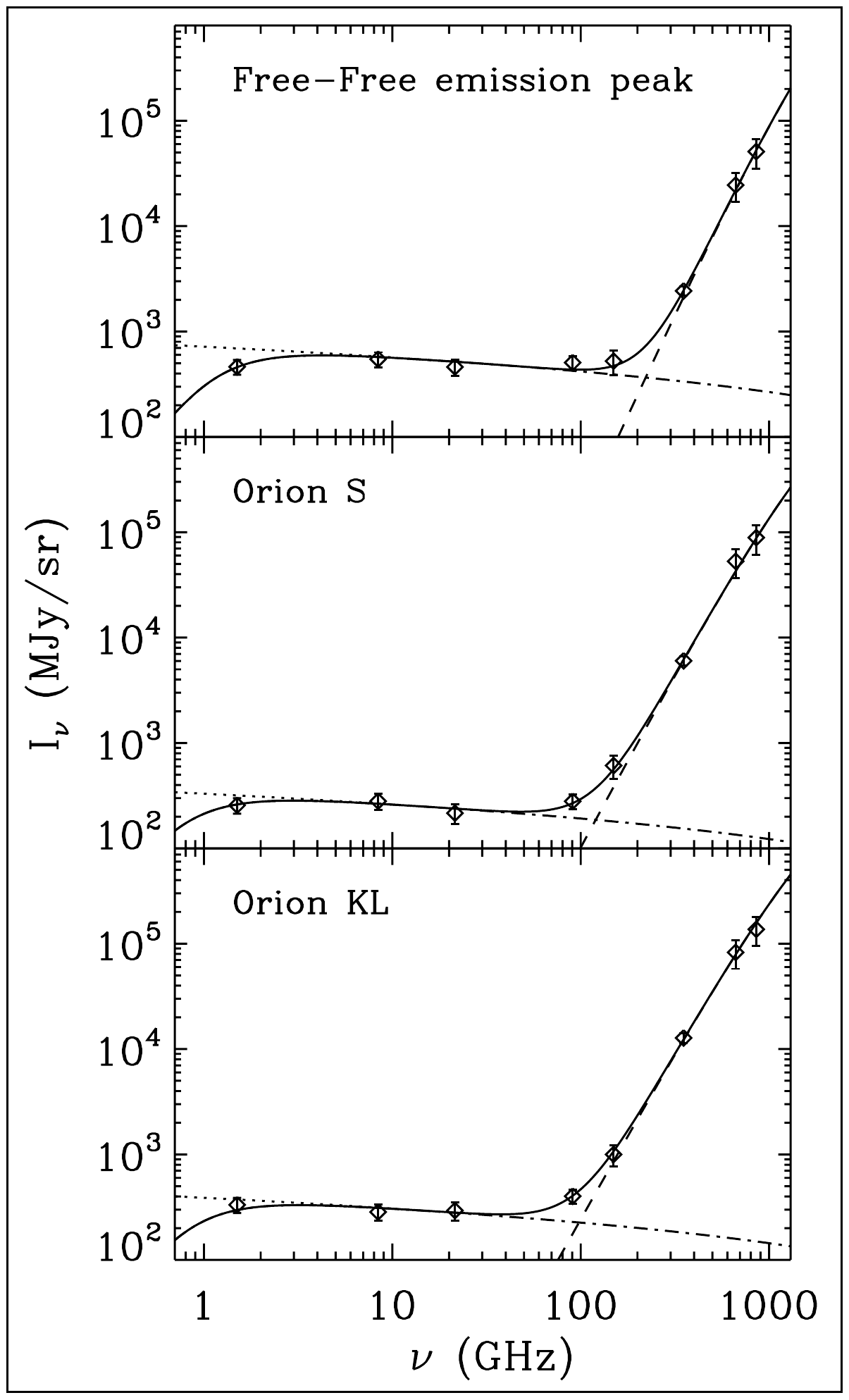} 
    \caption{ Average spectra over the areas delineated by the 30$''$
      diameter circles on Figure\,\ref{fig:imaffi_all}.  The dot-dashed
      line is the free-free emission component, the dashed line is the
      thermal dust emission component and solid line is the total. For
      comparison, the dotted line shows the free-free emission without
      taking account of the opacity.  This fitted function is
      described in the Appendix. }
         \label{fig:spectra}
\end{figure}

To interpret the MUSTANG data we made comparisons to maps at
other frequencies. These included a 854\,GHz (350\,$\mu$m) SHARC-II
map
and the 664\,GHz (450\,$\mu$m) and 352\,GHz
(850\,$\mu$m) maps from SCUBA \citep{johnstone99}.  We obtained new
150\,GHz maps using data from GISMO \citep{GISMO2008} and radio
frequency data at 21.5\,GHz, 8.4\,GHz, and 1.5\,GHz ($K$ band, $X$ band, and
$L$ band) from the GBT and the
VLA.  A summary of the angular resolution, RMS, and calibration
uncertainties of each of these maps is given in
Table~\ref{tab:datasets}.  In order to compare them, all the maps were
converted to MJy\,sr$^{-1}$ by dividing by the solid angles of their 
respective beams.

\subsection{The 1.5\,GHz map}

The image of M42 at 1.5\,GHz (20\,cm) was derived from archival VLA data
taken with the VLA in ``B'', ``C'', and ``D'' configurations.  This
combination of configurations gives sensitivity to angular scales
between 14$'$ and 4$''$.  Initial calibration of the data were carried
out in AIPS\footnote{http://www.aips.nrao.edu}
 using standard VLA flux density calibrators (3C48, 3C138, and 3C2860)
 and B0500$+$019 as an astrometric calibrator.

After the individual observations were calibrated, they were
concatenated and processed in Obit.  Automated flagging of RFI was
followed by an imaging using the ``SDI'' technique \citep{SDI}.
This technique uses a standard
visibility-based CLEAN to the point that the fine scale structure has
been removed from the residual and only a large, flat plateau is left.
In subsequent iterations, a CLEAN component is added at the
location of each residual pixel above a given limit.  This 
suppresses instabilities inherent with CLEAN that can produce artifacts for
very extended sources. Twenty one facets were used to cover the
full field of view as well as outlying sources. Twenty million CLEAN
components with a total flux density of 324\,Jy were used in the
deconvolution.  A Briggs Robust weighting parameter of -0.5 gave an
approximately 7$''$ FWHM beam and the restoration used a 7$''$
FWHM Gaussian beam.

\subsection{The 8.4\,GHz Map}

The 8.4\,GHz image was made by combining continuum observations from the GBT
and the VLA. Data from the GBT was taken in an
``On-The-Fly'' raster mode and contains information on angular
scales from the size of the map down to the limit of the GBT's resolution
at 8.4\,GHz (1.46$'$). The VLA data, taken with the
array in ``D'' configuration, are a series of tiled $3' \times 3'$
pointings and contain information on spatial scales from those of the
shortest to the longest baselines (2$'$ to 8.4$''$).  These data sets
were combined using AIPS++ and CASA\footnote{Information on CASA - The Common
  Astronomy Software Applications package is available at
  http://science.nrao.edu/index.php/Data\%20Processing/CASA}  to
produce a single map with a very high dynamic range and sensitivity to
all structures down to a FWHM resolution limit of 8.4$''$.
 
\subsection{The 21.5\,GHz Map}

A 21.5\,GHz continuum map of M42 was obtained from archival GBT
data and reduced in IDL. The data were collected using the GBT's
Digital Continuum Receiver (DCR), a general-purpose continuum back
end. An internal noise
calibration source with a known noise temperature was pulsed once
within each integration and the data were calibrated to antenna
temperature using this noise source. Data were collected with simple,
cross-linked raster scans 12$'$ long in Right Ascension and Declination. A
DC offset and a gradient were removed from each scan, and the
resulting data gridded onto the sky. The resulting map contains 
information on angular scales from 6$'$ to 33.5$''$, the resolution of
the GBT at 21.5\,GHz.

\subsection{The 150\,GHz GISMO Map}

GISMO is another TES bolometer camera built to operate at 150\,GHz on
the 30\,m IRAM telescope \citep{GISMO2008}.  
The data presented in this paper were obtained during the first run
with the instrument in November 2007. A ``data cleaning'' routine was
used to remove spikes in the data and to identify bad pixels. Since
the weather was exceptionally good ($\tau_{225}=0.07$) a high-pass filter
with a low frequency cutoff at 0.2\,Hz removed most of the
atmospheric noise. After this step a visual inspection of the data of
each individual pixel was carried out and pixels with spikes or steps
that were not already identified by software were flagged as
bad. Per-pixel gains and offsets were then calculated by fitting a
linear function to the correlation between the time series of data for
each single pixel and the time series of median values for each
frame. Pointing and flux calibrations were obtained by observations of
quasars and planets. The total integration time for the part of the
map shown in this paper was about 4 minutes. A combination of fast
scanning and GISMO's $2'\times 4'$ instantaneous field of view ensured
that all extended structure up to (and probably beyond) an angular size
of 2$'$ was recovered.

\subsection{The SCUBA Maps}
We used maps of the Orion ridge originally observed by
\citet{johnstone99}.
The data were taken of a 50$'$ long region over four nights with the
SCUBA instrument on JCMT.  The observations consist of a number of
10$'$ by 10$'$ maps made with chop throws of 20$''$, 30$''$, and
65$''$.  By combining maps made with these different chop throws and
different chop directions all extended structures with angular 
sizes up to the largest chop size (65$''$) were recovered.

\subsection{The SHARC-II Map}

We used the SHARC-II camera \citep{Dowell2003} on the Caltech
Submillimeter Observatory to obtain a 350$\mu m$
(854\,GHz) map.  
The absolute uncertainty for this measurement is  $\sim\,30\%$ and the 
beam size is 9$''$.  SHARC-II's field of view is $2.59' \times
0.97'$ so it can be expected that all extended structure with angular
scales less than 2$'$ is included in our map.

\section{THE MORPHOLOGY OF M42-OMC1}\label{sec:morphology}

Figure\,\ref{fig:imaffi_all} displays maps of the eight datasets we used.
Each dataset is presented at its nominal resolution.  At 21.5, 8.4
and 1.5\,GHz ($K$ band, $X$ band, $L$ band), 
the observed intensity is entirely dominated by
free-free emission from the ionized gas around the OB stars of the
Trapezium, since there is no evidence for significant synchrotron
emission in M42 as noted by \citet{subrahmanyan2001}.  
At higher frequencies, the SHARC-II (854\,GHz) and SCUBA (664 and 352\,GHz)
maps are dominated by thermal dust emission of the OMC1
molecular cloud lying behind the ionized gas. The two brightest
features seen with the thermal dust emission are Orion\,KL/BN and
Orion\,S\@.  
At 90 GHz, free-free and thermal dust emission have about
the same intensity in this region, and so both the ionized gas and the
OMC1 molecular cloud are seen in the MUSTANG map.
As the spectral shape of the dust emission is quite steep
compared to free-free emission, this equipartition vanishes
rapidly when looking at a different frequency. Thus, the spatial
structure in the GISMO map
 is dominated by the OMC1 features with only a relatively
low contribution from ionized gas.

To the southeast of the Trapezium stars is the well known ``Orion
bar'' associated with the edge-on boundary between the
H{\footnotesize\,II} region and the molecular cloud. A close up of the
bar is shown in Figure~\ref{fig:OrionBar}. The clear shift in the
position of the bar at different frequencies is what one would expect
at the ionization front.  Moving away from the exciting stars, one
sees first the ionization front emitting strongly at 90\,GHz
due to high emission measure (EM), next comes the mid-infrared
emission bands of polycyclic aromatic hydrocarbons from the
photodissociation region seen with IRAC 8\,$\mu$m \citep{IRACmap} and
finally, the molecular material which is bright at the SCUBA
352\,GHz (850\,$\mu m$) frequency due to the high column density of
dust.  Note that another much fainter bar can be seen in the MUSTANG
map to the north-east of the Trapezium at
$\alpha_{2000}\sim5^h\,35'\,23''$ and $\delta_{2000}\sim-5^{\circ}
\,20'\,20''$ (Figure\,\ref{fig:imaffi_mustang}). 
 This bar has already been detected at radio
frequencies \citep{yusef90} and with molecular tracers
\citep{fuente96, rodriguez-franco98}.

The spectra in Figure\,\ref{fig:spectra} are those observed at the
free-free emission peak, and at the Orion~KL/BN and Orion~S locations.
These are the averaged spectra over the areas delineated by circles on
Figure\,\ref{fig:imaffi_all}.  Before averaging each map was
brought to the spatial resolution of the 21.5\,GHz
map (33.5$''$) by convolving it with a Gaussian of appropriate FWHM\@.
Spectra were fitted using the function described in the
Appendix.  As noted previously, 
 at the MUSTANG frequency (90\,GHz), the emission toward Orion
KL/BN has almost equal contributions from thermal dust and 
free-free emission. However, toward the free-free emission peak,
free-free emission dominates the MUSTANG intensity and even the GISMO
one at 150\,GHz.

\begin{table}[t]
\centering
 \caption{A comparison between our EM values and those of \citet{baldwin91}.}
  \label{tab:bald} 
  \begin{tabular}{c c c}
   \hline           
   \hline
location relative  &   Baldwin et al.  & this paper\\ 
to  $\theta^1\,Ori\,C$   &   (${\rm pc\,cm}^{-6}$)   &  (${\rm pc\,cm}^{-6}$)   \\
   \hline
30$''$ west &$8.8\times 10^6$&$8.2\pm1.1\times 10^6$\\  
50$''$ west &$4.3\times 10^6$&$3.8\pm0.5\times 10^6$\\  
100$''$ west&$8.8\times 10^5$&$5.3\pm0.7\times 10^5$\\  
   \hline
   \end{tabular}
\end{table}

\begin{figure}[!t]
   \centering
    \includegraphics[width=0.45\textwidth]{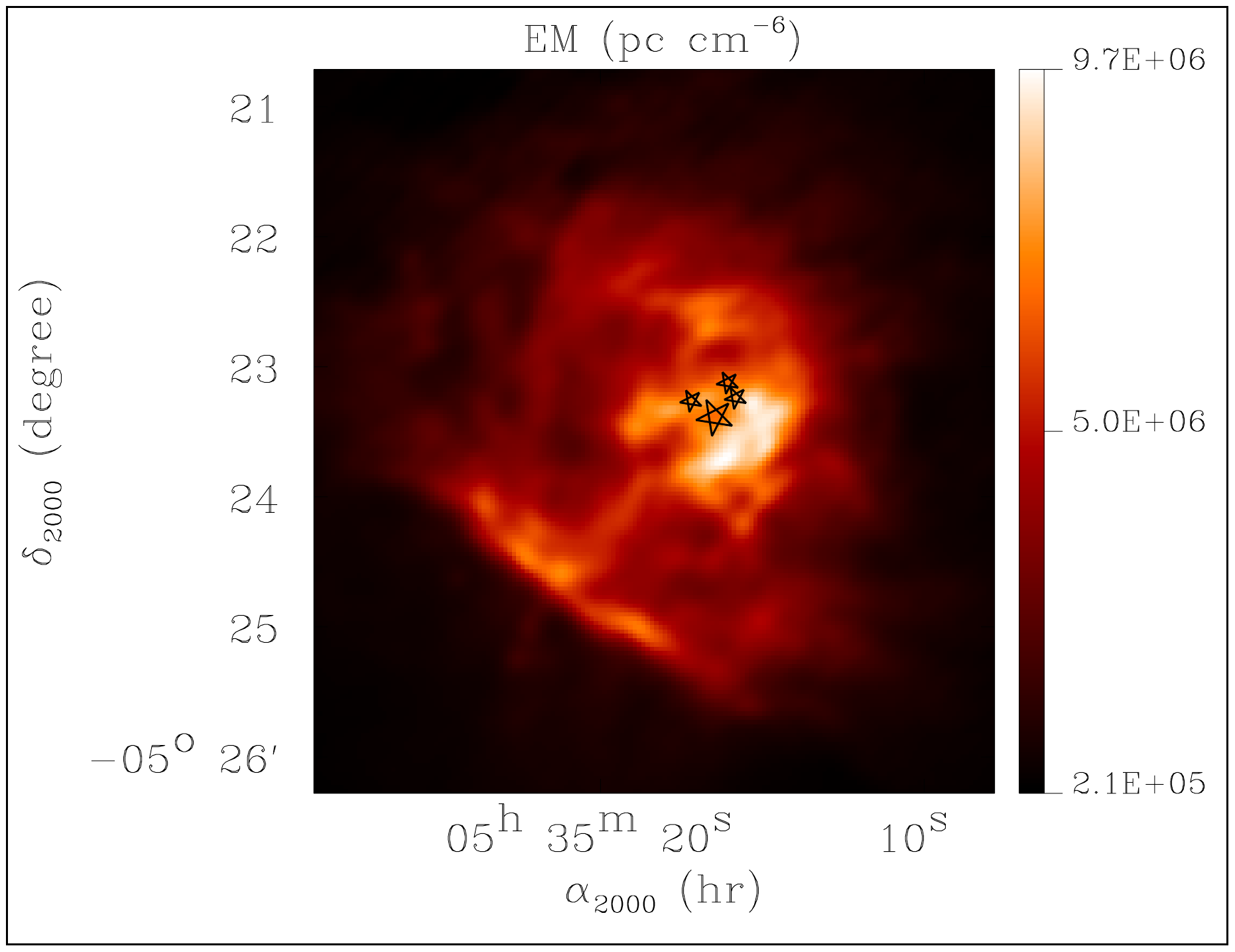}
    \caption{A map of the emission measure derived from the
    8.4\,GHz and 1.5\,GHz data at 8.4$''$ resolution.  The stars
    show the location of the Trapezium stars.}
         \label{fig:em}
\end{figure}

Toward the free-free emission peak, we obtained an emission measure
(EM) value of
$7.6\pm0.8\times 10^6\,{\rm pc\,cm}^{-6}$ in a 33.5$''$ FWHM
beam. This value can be compared with a value of $2.7\times 10^6\,{\rm
  pc\,cm}^{-6}$ in a 90$''$ FWHM beam at 10.55\,GHz by
\citet{subrahmanyan2001}. Taking into account the beam dilution from
33.5$''$ to 90$''$, we obtained a value of $5.1\pm0.5\times 10^6\,{\rm
  pc\,cm}^{-6}$, a factor $\sim1.9$ above the \citet{subrahmanyan2001}
value. This discrepancy could, in part, come from the fact
that they used a differencing receiver which could potentially miss
flux on spatial scales similar to the separation of the feeds.  We can
also compare our EM estimate with those of \citet{baldwin91} that were
deduced from H11-3 measurements.  We derived an EM map
(Figure~\ref{fig:em}) with 8.4$''$
resolution using only the 8.4\,GHz ($X$ band) and 1.5\,GHz ($L$ band) data which
can be directly compared with the East-West cut
toward the West of $\theta^1$\,Ori\,C given by \citet{baldwin91}.  As
can be seen in Table~\ref{tab:bald} both the absolute value and the
spatial variation of our radio estimate of EM are in good agreement
with their optical one.

We found that no anomalous emission component is needed to reproduce
data at all frequencies and angular scales ($<1.1'$) 
considered in this paper.  A number of
detections of such emission have been made in other parts of the sky, 
for example \citet{Kogut96} found an
excess between 30\,GHz and 90\,GHz at high latitudes in COBE data. This
excess is correlated with DIRBE far-infrared data and has since been
interpreted as arising from spinning dust
\citep[e.g., ][]{lazarian2003}. Excess emission on large spatial scales
has also been detected at frequencies of 15\,GHz and 30\,GHz by
\citet{Leitch97} and at 33\,GHz and 43\,GHz by \citet{Oliveira97}.
Anomalous emission was also observed toward individual objects such as
H\,{\scriptsize II} regions \citep[e.g., ][]{dickinson2006,
  dickinson2009}, cold dense clouds \citep[e.g., ][]{Casassus2006} and
planetary nebula \citep[e.g., ][]{Casassus2007}. Emission from
spinning dust is thought to peak between 10\,GHz and 50\,GHz so a
small component could be missed in our analysis. 
\citep{lazarian2003}.  In order to check for the presence of anomalous
emission, new data at $Ka$~band (27\,GHz--42\,GHz) have been obtained and
analysis is currently underway.

\begin{figure}[!t]
   \centering
    \includegraphics[width=.45\textwidth]{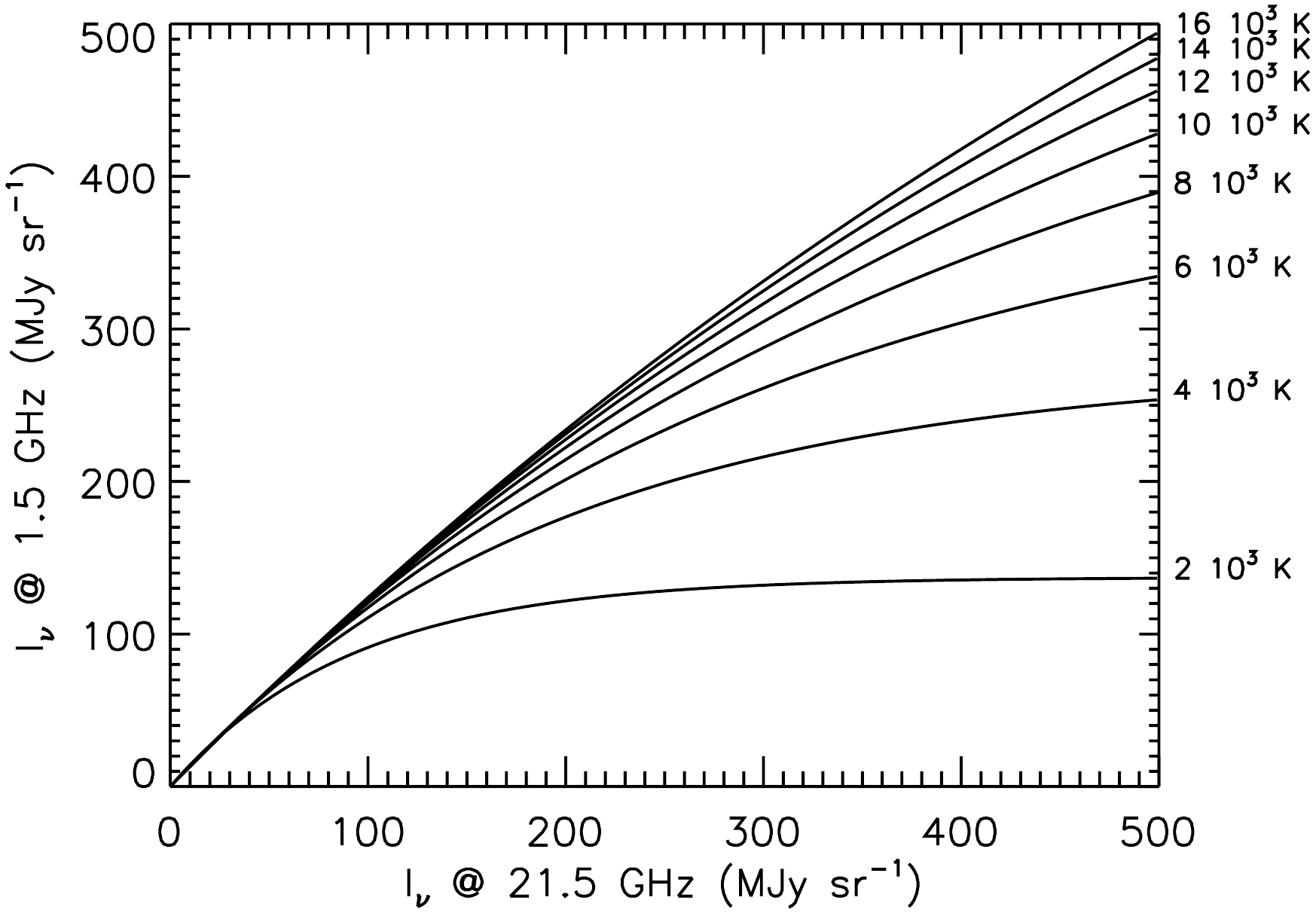} 
    \includegraphics[width=.40\textwidth]{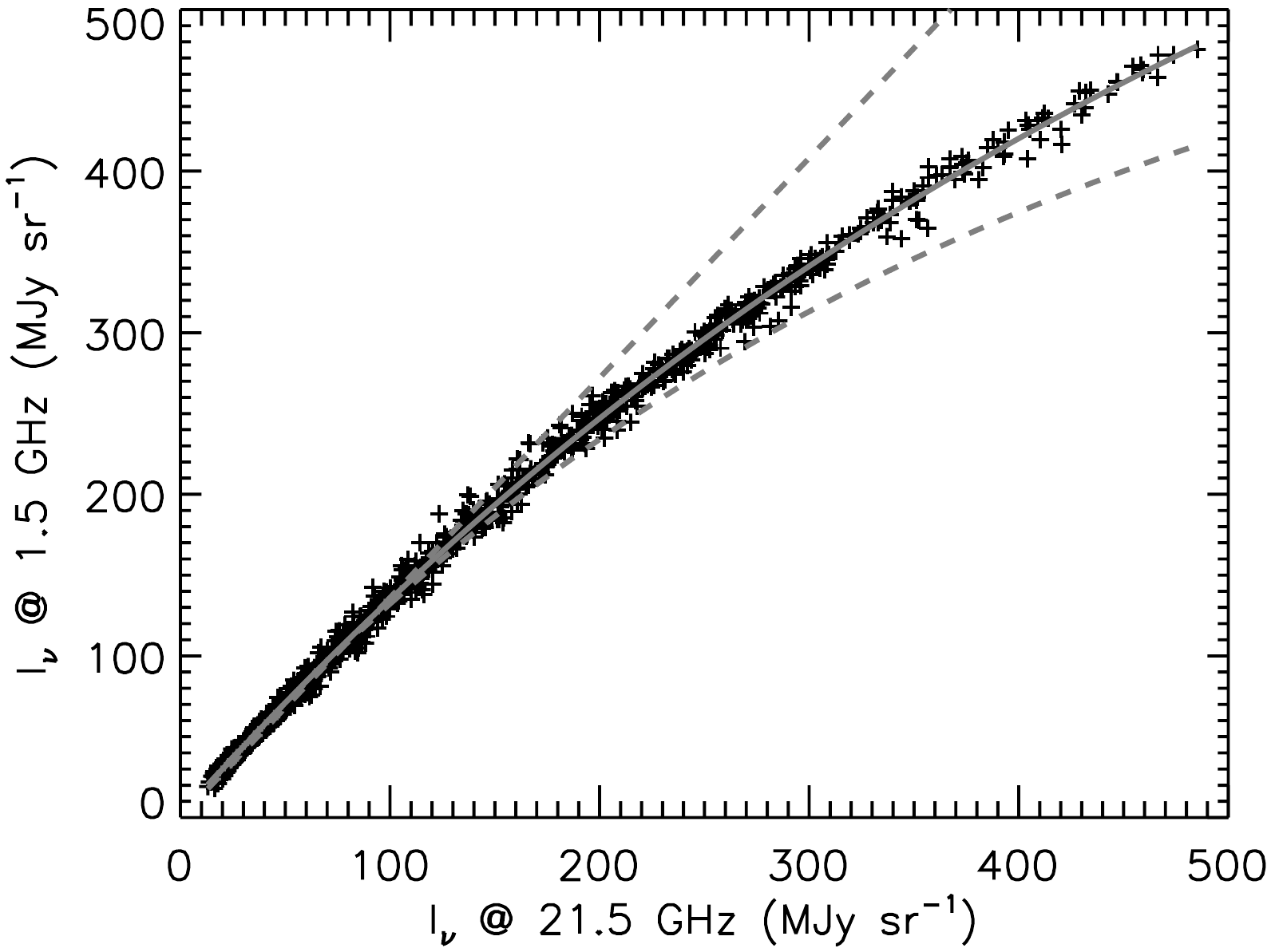} 
    \caption{Modeled (top panel) and observed (bottom panel)
      relationships between 1.5\,GHz and 21.5\,GHz free-free emission
      intensities. On the bottom panel, in gray, is the
      relationship for the fitted electron temperature of
      $T_e\,=\,11376$\, K\@.  For comparison, the curved dashed line is
      the expected relationship for $T_e\,=\,7865$\,K
      \citep{subrahmanyan92} and the straight dashed line is the
      expected relationship without opacity (i.e. $T_e\,=\,\infty$).}
         \label{fig:correl_LvsK}
\end{figure}

\begin{figure}[t!]
   \centering
    \includegraphics[width=.45\textwidth]{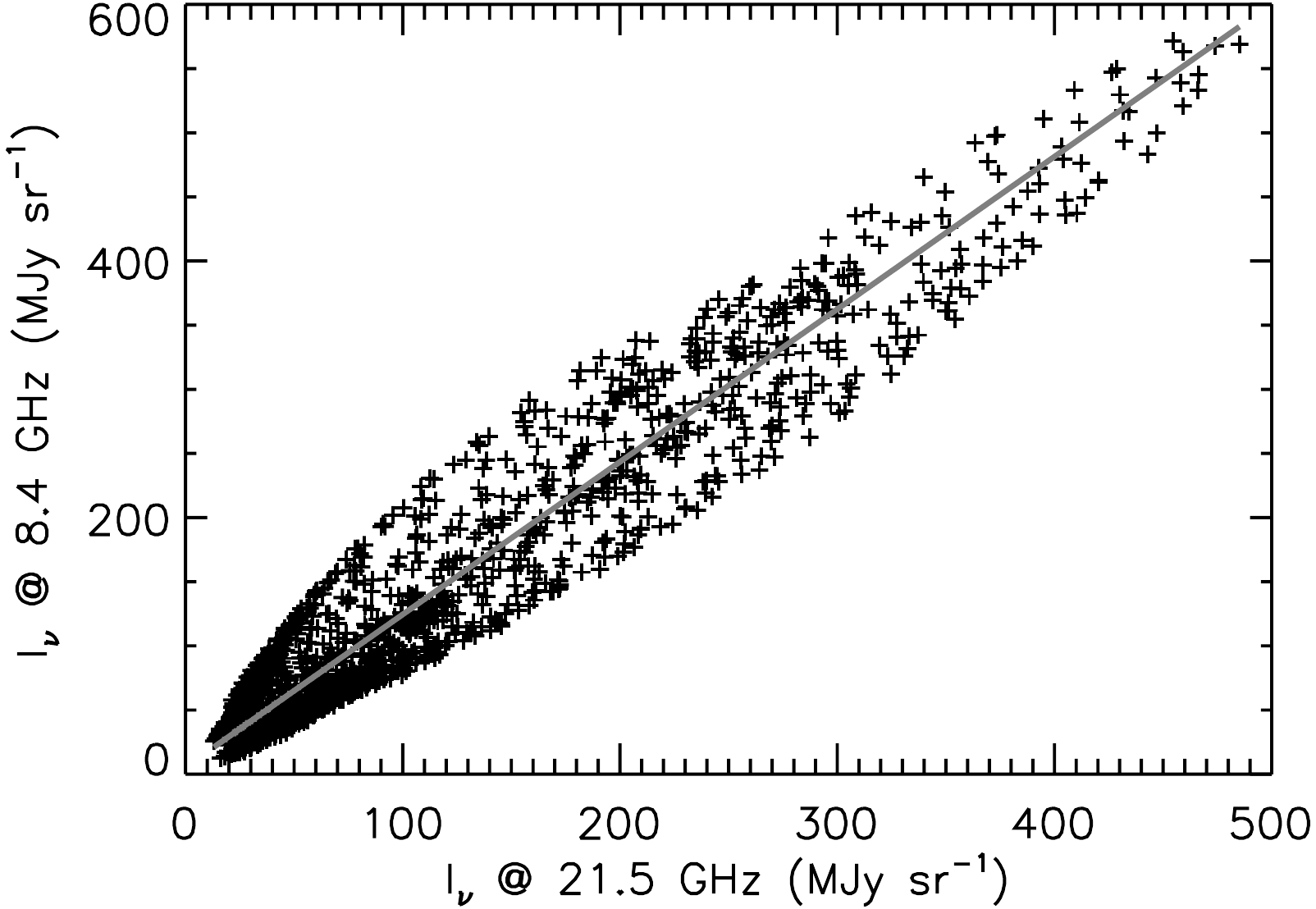} 
    \includegraphics[width=.45\textwidth]{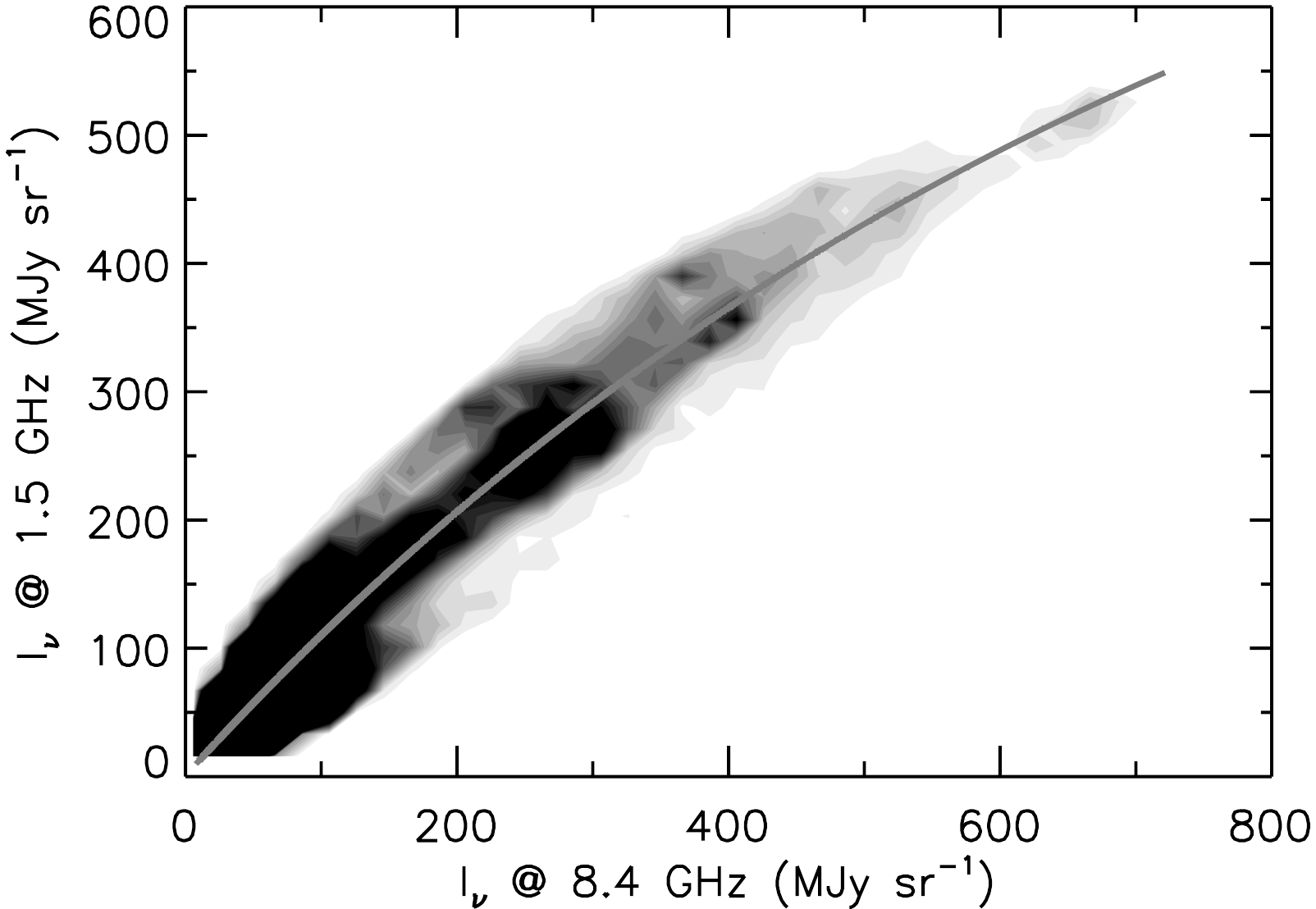} 
    \caption{Relationship between 21.5\,GHz and 8.4\,GHz (top panel) and
      8.4\,GHz and 1.5\,GHz (bottom panel) free-free emission
      intensities. For the bottom panel, we show the histogram of points density
     in gray-scale.
      Gray lines are the fitted relationship (see
      Section~\ref{sect:ff_correl}).  }
         \label{fig:correl_ff}
\end{figure}

\section{$T_e$ determination from the free-free intensity relationship}\label{sect:ff_correl} 
Assuming that free-free emission is optically thin at 
frequency $\nu_1$ and potentially optically thick at frequency
$\nu_2$, the relationship between intensities at these two
frequencies is
\begin{equation}{
I_{\nu_2}\,=\,I_{\nu_1}\,\frac{g_{ff}(T_e, \nu_2)}{g_{ff}(T_e,
  \nu_1)}\,\frac{{\rm e}^{-\frac{h\,\nu_2}{k\,T_e}}}
{{\rm e}^{-\frac{h\,\nu_1}{k\,T_e}}}\,\frac{1\,-\,{\rm e}^{-\tau(I_{\nu_1}, 
    T_e, \nu_2)}}{\tau(I_{\nu_1}, T_e, \nu_2)}} ,
\label{eq:ff_correl}
\end{equation}
with $g_{ff}(T_e, \nu_2)$ being the Gaunt factor defined by
Equation\,\ref{eq:gaunt_factor} and $\tau(I_{\nu_1}, T_e, \nu_2)$ the
optical depth as defined by Equation\,\ref{eq:tau} (with
$EM\,\propto\,I_{\nu_1}$).  The top panel of
Figure~\ref{fig:correl_LvsK} shows this function for 1.5\,GHz
($I_{\nu_2}$) versus 21.5\,GHz ($I_{\nu_1}$) at different electron
temperatures. Fitting this function to the observed relationship
provides $T_e$, formally the only independent parameter of
Equation\,\ref{eq:ff_correl}.  To take into account
possible calibration issues between the datasets we added a
multiplicative parameter $b$ which corresponds to the gain ratio
and an additive parameter $a$, that compensates for
possible DC offsets between the two datasets.  We then fitted
$I_{\nu_2}\,=\,b\,\times\,f(T_e)\,+\,a$ with $f(T_e)$ given by
Equation\,\ref{eq:ff_correl}.
The bottom panel of Figure\,\ref{fig:correl_LvsK} shows the observed
1.5\,GHz ($I_{\nu_2}$) vs 21.5\,GHz ($I_{\nu_1}$) relationship together
with the modeled function for the best fit $T_e\,=\,11376\pm1050$\,K\@.  
Note that in addition to the very low scatter about the
observed relationship, the gain ratio (b) and DC offset (a) between
the two datasets is $1.08\pm0.02$ and $1.4\pm1.0$\,MJy\,sr$^{-1}$,
respectively, which lie within the calibration uncertainties given in
Table~\ref{tab:datasets}.

As shown in Figure~\ref{fig:correl_ff}, the relationship between free-free
intensities involving 8.4\,GHz ($X$-band) data are noticeably scattered.
This could be related to a flux calibration discrepancy between the GBT
and VLA 8.4\,GHz data.  The overall relationship for 8.4\,GHz vs
21.5\,GHz intensities is linear as expected between two optically thin
frequencies.  Shown on this plot is the line with a fitted slope of
1.19$\pm$0.12. The predicted slope is 1.12 (assuming our
$T_e\,=\,11376$\,K estimate).  As the slope depends only weakly on
$T_e$, the difference between the measured and the expected values is plausibly
related to the gain ratio between the two datasets and is consistent
with calibration uncertainties given in Table~\ref{tab:datasets}.  The
1.5\,GHz vs 8.4\,GHz intensity relationship (bottom panel of
Figure~\ref{fig:correl_ff}) is non linear as predicted by
Eq.~\ref{eq:ff_correl}.  The best fit electron temperature in this
case is $T_e\,=\,13457\pm3000$\,K, in good agreement with that from
Figure\,\ref{fig:correl_LvsK}.  Due to the high scatter of the 1.5\,GHz
vs 8.4\,GHz intensity relationship, we adopt the former
($T_e\,=\,11376\pm1050$\,K).

This estimate is significantly above the  $7865\pm360$\,K given by
\citet{subrahmanyan92} that was estimated directly
from the brightness temperature at the free-free emission peak in a
330\,MHz map, where the emission should be optically thick.
In M42 most of the ionized gas emission comes from the main ionizing
front (MIF) located beyond $\theta^1$\,Ori\,C at the boundary with
OMC1. This MIF is seen face-on.  Therefore, when there is significant
free-free optical depth there is a ``photosphere'' soon reached on the
near side of the MIF\@.  This is certainly the case for the measurement
by \citet{subrahmanyan92}.
It is well established that the electron temperature rises from the
ionizing star to the ionizing front, due to the hardening of the
radiation field as the lowest energy photons are absorbed first. 
Using different ions emitting at successively greater depths,
\citet{Esteban98} has observed the increase of $T_e$ as one approaches
the MIF on a single line of sight:
$T\,=\,8730\pm800$\,K from the Balmer lines-to-continuum ratio,
$T_e[\mbox{O\,{\scriptsize III}}]\,=\,8300\pm210$\,K,
$T_e[\mbox{N\,{\scriptsize II}}]\,=\,9850\pm375$\,K and
$T_e[\mbox{S\,{\scriptsize III}}])\,=\,10300^{+2440}_{-960}$\,K.
Moreover, according to the modeling of the MIF with the software
package Cloudy \citep[08.00, latest
described in][]{ferland98}, the 330\,MHz photosphere is located at the
same depth into the H\,{\scriptsize II} region as the
[O\,{\scriptsize III}] emission.  Since the 
opacity at 1.5\,GHz ($L$-band) is $\tau\la1$ over most of the map, with
our correlation method, we estimated the electron temperature weighted
by the EM\@.  The highest densities and hence the highest opacities
are at the neutral edge of the MIF so our estimate of  $T_e$ is biased
toward this region.

\section{Dust emission}\label{sect:dust}

\begin{figure}[t!]
   \centering
    \includegraphics[width=0.45\textwidth]{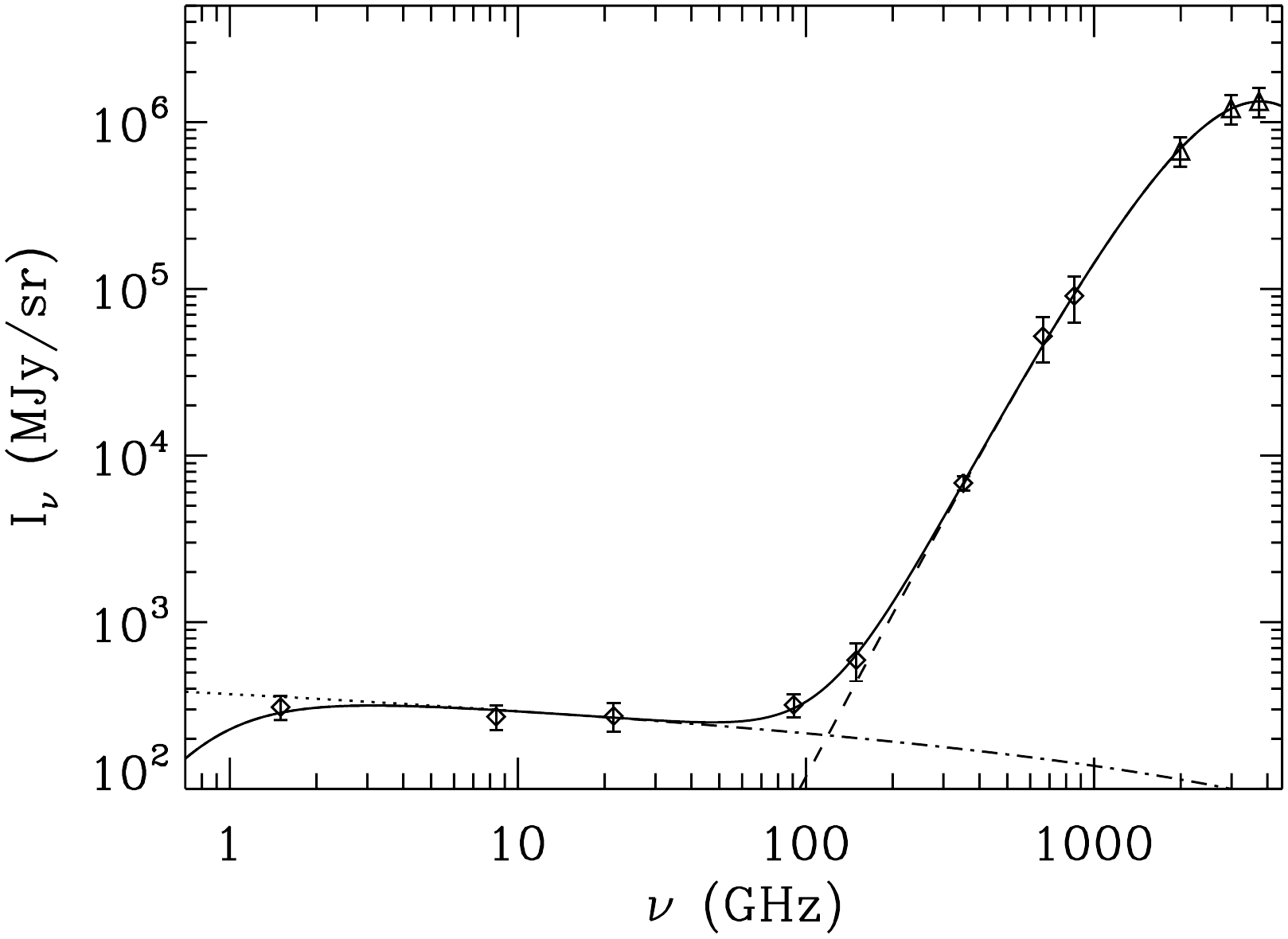} 
       \caption{Averaged spectrum of Orion\,KL/BN in the 80$''$ beam of 
         {\it ISO} LWS.
                      Triangles are ISO LWS data taken from \citet{lerate2006}. 
                      The dot-dashed line is the free-free
                      emission component, 
                      the dashed line is the thermal dust emission component
                      and solid line is the total. For comparison, the dotted line
                      shows the free-free emission without taking into account the opacity.
                      This fitted function is described in the Appendix.
                      }
         \label{fig:spectre_KL_LWS}
\end{figure}

Toward M42-OMC1, dust emission from (1) the cold dense part of OMC1,
(2) the hot interface of the molecular cloud heated by the Trapezium
stars, and (3) the H\,{\scriptsize II} region, are mixed.  
In their study based on PRONAOS submillimeter observations and
ancillary data as short as 200 and 90 \micron, \citet{dupac2001}
derived a temperature of 66\,K for the dust.
Hot dust fills a larger fraction of 3.5$'$ PRONAOS beam than cold
dust. Consequently, even if only from
the H\,{\scriptsize II} region and the surface of OMC1 (which has a
relatively small column density compared to that of the cold component
extending through the molecular cloud), emission from hot dust is
energetically important.  This is the situation
expected as the depth of the hot dust in OMC1 is set by when the
opacity to UV light, $\tau_{UV}$, is approximately 1.
Moreover, for the shortest wavelengths they used 
(90 and 200\,$\mu$m), where the hot dust dominates the emission.  
\citet{werner76} derived a dust temperature ranging between 55
and 85\,K from the 50/100$\,\mu$m emission ratio in a 1$'$ beam.

At smaller angular scales toward the regions of the highest column density
such as Orion\,KL, the colder dust component from the more shielded part
of the molecular cloud must dominate. Especially at wavelengths longer
than 350\,$\mu$m, we expect to find a lower apparent dust temperature (even
allowing for embedded energy sources such as BN and Irc1).
Figure~\ref{fig:spectre_KL_LWS} displays the averaged spectrum of
Orion\,KL/BN in the 80$''$ beam of the {\it Infrared Space Observatory} long wavelength spectrometer ({\it ISO} LWS), using our dataset and three
points at 80, 100 and 150\,$\mu$m taken from the {\it ISO} LWS data of
\citet{lerate2006}.  Fitting the function described in the
appendix, we obtained a dust temperature
$T_d\,=\,42\pm3$\,K and a spectral emissivity index
$\beta_d\,=\,1.3\pm0.1$ for the dust, lower than the average over the
nebula.
\citet{lis98} derived a value of $\beta_d\,=\,1.8$ toward Orion\,KL/BN
from the 350/1100\,$\mu$m emission ratio and assuming a temperature of
55\,K\@.  They would have obtained an even higher value of $\beta_d$ by
using $T_d\,=\,42$\,K.  This difference could be related to their use
of the 3300$\,\mu$m (90\,GHz) data of \citet{salter89} to subtract
the free-free contribution from their 1100\,$\mu$m data.  As we
discussed in Section~\ref{sec:morphology} and is evident in
Figure~\ref{fig:spectre_KL_LWS}, dust emits significantly at
3300$\,\mu$m (90\,GHz) toward Orion\,KL/BN so that they must have
overestimated the free-free contribution at 1100$\,\mu$m leading to an
underestimation of the dust emission and then to an overestimation of
$\beta_d$ using the 350/1100\,$\mu$m emission ratio.

Our dataset, summarized in Table~\ref{tab:datasets}, does not go out to 
short enough wavelengths to allow us to
constrain both $\beta_d$ and $T_d$ at other points. 
Therefore, to study $\beta_d$ variations between
the free-free emission peak, Orion\,S and Orion\,KL/BN locations
(shown on Figure~\ref{fig:imaffi_all}) we fixed $T_d$ to the value of
42\,K derived above.
Spectra for these positions are shown in Figure~\ref{fig:spectra} and
were obtained by averaging the signal within a 30$''$ diameter circle.
We derived values of $\beta_d\sim\,2.0\pm0.1, 1.4\pm0.1$ and
$1.2\pm0.1$ for the free-free emission peaks, Orion\,S and Orion\,KL/BN,
respectively.  At the Orion\,KL/BN location, $\beta_d$ goes from 1.3
to 1.2 as the area averaged over is decreased from 80$''$ to 30$''$ 
in diameter.  The decrease of $\beta_d$ toward denser regions of OMC1
has been reported previously \citep[e.g.,
][]{goldsmith97,lis98, johnstone99}. This may not be a physical
effect, but rather the result of describing the spectral energy distribution
 from dust that has a
distribution of temperatures using a single value of $T_d$.

In order to study the trend of $T_d$ related to the observed spectral
evolution, we fitted the same three spectra with fixed $\beta_d\,=\,1.8$ (the
low value of $\beta_d\,=\,1.3$ obtained toward Orion\,KL/BN would give an
unphysically high temperature for the free-free emission peak).  We got
$T_d\,=\,82\pm10, 21\pm1$ and $15\pm1$\,K for the free-free emission peaks,
Orion\,S and Orion\,KL/BN, respectively.  One must note that the
obtained temperature at the free-free emission peak is then close to
the maximum value $T_d\,\sim\,85$\,K from \citet{werner76}.  Moreover,
$T_d\,\sim\,82\,K$ and $\beta_d\sim\,1.8$ lead to peak emission at
$\lambda\,=\,30-40\,\mu m$. This is consistent with the prediction of
the dust model described in \citet{compiegne2008} with an intensity of
the UV radiation field of $\chi\,\sim10^4$, compatible with the
expected value at the hot interface of the molecular cloud heated by the Trapezium
stars ($\chi\,\lesssim\,10^5$).
Toward Orion\,KL/BN, $T_d\,\sim\,15$\,K is a lower limit 
($\beta_d\,\sim\,1.3$ from the spectrum of
Figure~\ref{fig:spectre_KL_LWS} leads to $T_d\,\sim\,30$\,K) but
is nevertheless feasible for dust in a dense shielded environment.

Finally, it is worth emphasizing that a spectral dependence
on evolution of both temperature and spectral
emissivity index of the dust is plausible. 
We have already mentioned a decrease in temperature
from $\sim$\,85\,K to $\sim$\,20\,K going from
the highly excited environment around the Trapezium (the free-free
emission peak) to the denser regions shielded from the UV
radiation field (Orion\,KL/BN and Orion\,S spectra).  
A decrease of the spectral emissivity index toward
a denser medium can arise from dust grain properties evolution (shape,
size, porosity, composition) following processes such as coagulation
or accretion.  
It is therefore remarkable that although the observed spectra
are averages along the lines of sight among different dust emission
properties, a single dust temperature and spectral emissivity index
suffice to reproduce the spectrum between 350 and 3300\,$\mu$m.
\citet{goldsmith97} also reported no systematic trend indicative of
frequency dependence of $\beta_d$ between 450\,$\mu$m and 1100\,$\mu$m
toward this region.

\section{Conclusion}\label{conc}

We have presented one of the first maps made with the MUSTANG camera that
allows observations at 90\,GHz using the 100\,m diameter Green Bank
Telescope (GBT).  We obtained a $5' \times 9'$ continuum map in the
bright Huygens region of M42 at 9$''$ resolution with a
sensitivity of 2.8\,mJy/beam.  Dual data analysis pipelines produced
similar maps which are free from artifacts and have dynamic ranges of
over 1000.

To analyze the physics of this region, we used
 a multi-frequency dataset ranging from 1.5\,GHz to 854\,GHz whose
characteristics are summarized in
Table\,\ref{tab:datasets}.  We also presented a newly obtained 150\,GHz
map from the GISMO camera on the 30\,m IRAM telescope. 
90\,GHz is an interesting transition frequency, such that MUSTANG
detects both the free-free emission characteristic of the
H\,{\scriptsize II} region created by the Trapezium stars, normally
seen at lower frequencies, and thermal dust emission from the
background OMC1 molecular cloud, normally mapped at higher
frequencies.  Using an analytical spectral function (described in
the appendix)  including free-free and thermal
dust emission, we studied these two components.

We derived a peak value of the emission measure (EM) of
$7.6\pm0.8\times 10^6\,{\rm pc\,cm}^{-6}$ in a 33.5$''$ FWHM beam (the
resolution of our 21.5\,GHz map).  Using only higher resolution radio
frequency maps ($\sim\,8.4''$), we were able to get the spatial
variations of the EM at the same resolution as optical estimates by
\citet{baldwin91}.  Both the absolute value and the spatial variation
of our EM estimate are in close agreement with those of
\citet{baldwin91}.

We derived a new estimate of the EM averaged
electron temperature of $T_e\,=\,11376\pm1050\,K$. Our method is
based on the fitting of the non-linear behavior of the relationship of
free-free emission intensities at optically thin and optically thick
frequencies whose only free parameter is the electron temperature.  Our
temperature estimate is in agreement with previous estimates for the
main ionization front $T_e([\mbox{S\,{\scriptsize III}}])\,=\,10300^{+2440}_{-960}$\,K \citep[see][]{Esteban98} 
where the density rises and is then expected to dominate the EM.

We used {\it ISO}-LWS data in addition to our dataset in order to
constrain both the temperature and spectral emissivity index of the
dust within the 80$''$ {\it ISO}-LWS beam toward Orion\,KL/BN.  We derived
$T_d\,=\,42\pm3\,K$ and $\beta_d\,=\,1.3\pm0.1$.  
By comparing spectra from the free-free emission peaks,
Orion\,S and Orion\,KL/BN we have shown that $T_d$ and $\beta_d$ decrease
 from the H\,{\scriptsize II} region and the excited OMC1 interface
to the denser UV shielded part of OMC1 (Orion\,KL/BN, Orion\,S).
While the temperature tendency is obviously related to the UV exciting radiation
field and its extinction in the molecular cloud, the  $\beta_d$ tendency could be related
to dust grain properties evolution (shape, size,
porosity, composition) following processes such as coagulation or
accretion. Since the MUSTANG data presented in this paper was taken
there have been significant improvements to MUSTANG's sensitivity and
more are expected in the future.  MUSTANG is available to the
community and we expect a bright future for 90\,GHz science using the GBT.

The authors would like to acknowledge all those who worked to make
the MUSTANG camera and we would also like to thank the GISMO team and
the IRAM staff for
their hard work. We thank Rick Arendt
and Dale Fixsen for sharing their least squares imaging code and
Darren Dowell for his work on the SHARC-II map.
 The GISMO observations were made possible
by support through NSF grant AST-0705185.  Funding for the MUSTANG camera was
provided by National Radio Astronomy Observatory (NRAO) and the
University of Pennsylvania.
Observations were supported by NSF award number AST-0607654. The
National Radio Astronomy Observatory
 is operated by Associated Universities Inc. under a cooperative
 agreement with the National Science Foundation.

\appendix

\section{Fitted spectral function}\label{sect:spectral_fit}

The function used to fit the spectra is
\begin{equation}
I_{\nu, fit}\,=\,I_{\nu, d}+\,I_{\nu, ff} ,
\end{equation}
with $I_{\nu, d}$ the thermal dust emission intensity and $I_{\nu,
  ff}$ the free-free emission intensity. 
We have
\begin{equation}
I_{\nu, d}\,=\,C_{d}\,~\,B_{\nu}(T_{d}, \nu)\,~\,\nu^{\beta_{d}} ,
\end{equation}
with $\nu$ the frequency, $C_{d}$ a parameter relating to the dust
opacity and column density, $B_{\nu}(T_{d}, \nu)$ the Planck function
at the temperature $T_{d}$ of the dust, and $\beta_{d}$ the spectral
index for the dust emissivity.  The free-free emission intensity is
expressed as
\begin{equation}
I_{\nu, ff}\,=\,I_{0, \nu, ff}(EM, T_e,
\nu)\,\times\,\frac{(1\,-\,e^{-\tau(EM, T_e, \nu)})}{\tau(EM, T_e,
  \nu)} 
\end{equation}
with 
\begin{equation}
I_{0, \nu, ff}(EM, T_e, \nu)\,=\,\,C_{ff}\,\,T_e^{-1/2}\,\,g_{ff}(T_e,
\nu)\,\,e^{-\frac{h\nu}{kT_e}}\,\,EM 
\end{equation}
being the intensity without opacity at
frequency $\nu$ for 
a value EM of the emission measure, an electron temperature $T_e$ and with 
$C_{ff}\,=\,5.4\,10^{-39}$
taking the mean ion charge to be $\sim\,1$ and the refractive index of the 
plasma  $\sim\,1$.
We use the Gaunt factor approximation
\begin{equation}
g_{ff}(T_e, \nu)\,=\,\frac{\sqrt{3}}{\pi}\,ln(5\times10^7\,\frac{T_e^{-3/2}}{\nu}).
\label{eq:gaunt_factor}
\end{equation}
The optical depth is defined as
\begin{equation}
\tau(EM, T_e,
\nu)\,=\,8.235\times10^{-2}\,(\frac{T_e}{\mbox{K}})^{-1.35}\,(\frac{\nu}{\mbox{GHz}})^{-2.1}\,(\frac{EM}{{\rm
    pc}\,{\rm cm}^{-6}})\,a(\nu, 
T_e) 
\label{eq:tau}
\end{equation}
with $a(\nu, T_e)$ the correction factor tabulated by \citet{Mezger67}
that is close to 1.  
Among the 5 parameters ($C_{d}$, $T_d$, $\beta_d$, $EM$, $T_e$)  to
fit, we fixed 
$T_e$ to 11376\,K, the mean value that we derive in
Section~\ref{sect:ff_correl} (the dust parameters are not very
sensitive to this choice). The fitting was done using MPFIT
\citep{Markwardt2009}.

\bibliographystyle{apj}
\bibliography{orionPaper}

\end{document}